\documentclass[conference]{IEEEtran}
\IEEEoverridecommandlockouts
\usepackage{cite}
\usepackage{amsmath,amssymb,amsfonts}
\usepackage{xspace}
\usepackage{subfiles}
\usepackage{booktabs}
\usepackage{subfiles}
\usepackage{graphicx}
\usepackage{multicol}
\usepackage{subcaption}
\usepackage{array}
\usepackage[wby]{callouts}
\usepackage{url}
\usepackage{mathtools}
\usepackage{chngcntr}
\usepackage{algorithm}
\usepackage{titlesec}
\usepackage{framed}
\usepackage{tcolorbox}
\usepackage{adjustbox}
\usepackage{hyperref} 
\usepackage{array, booktabs, makecell}
\usepackage{siunitx, mhchem}
\usepackage{csquotes}
\usepackage{graphicx}
\usepackage{textcomp}
\usepackage{xcolor}
\usepackage{bm,amsmath}
\usepackage{amsfonts} 
\usepackage{physics}
\usepackage{multirow}
\usepackage{algorithm}
 \usepackage{algpseudocode}
 \usepackage[super]{nth}
 \usepackage[a4paper,
            bindingoffset=0.2in,
            left=1.7cm,
            right=1.7cm,
            top=1.9cm,
            bottom=2.7cm]{geometry}
 \usepackage{algorithmicx}
 \algdef{SE}[DOWHILE]{Do}{doWhile}{\algorithmicdo}[1]{\algorithmicwhile\ #1}%

\algrenewcommand\algorithmicrequire{\textbf{Input:}}
\algrenewcommand\algorithmicensure{\textbf{Output:}}
\makeatletter
\def\ps@IEEEtitlepagestyle{%
\def\@evenfoot{}%
}
\def\BibTeX{{\rm B\kern-.05em{\sc i\kern-.025em b}\kern-.08em
    T\kern-.1667em\lower.7ex\hbox{E}\kern-.125emX}}
\begin{document}

\title{Predicting System Crashes with Large Language Models
}


\author{\IEEEauthorblockN{\nth{1} Priyanka Mudgal*\thanks{*Previously at Intel while this work was conducted.}}
\IEEEauthorblockA{
\textit{Intel Corporation, USA}\\
priyanka.mudgal@intel.com}
\and
\IEEEauthorblockN{\nth{2} Bijan Arbab}
\IEEEauthorblockA{
\textit{Intel Corporation, USA}\\
bijan.arbab@intel.com}
\and
\IEEEauthorblockN{\nth{3} Swaathi Sampath Kumar}
\IEEEauthorblockA{
\textit{Intel Corporation, USA}\\
swaathi.sampath.kumar@intel.com}

}


\maketitle

\begin{abstract}
As the dependence on computer systems expands across various domains, focusing on personal, industrial, and large-scale applications, there arises a compelling need to enhance their reliability to sustain business operations seamlessly and ensure optimal user satisfaction. Achieving enhanced reliability depends on predicting and preemptively addressing these failures presents a formidable challenge. However, the challenge of predicting failures and mitigate them in timely manner is a complex problem. Thus, it becomes important to adopt effective strategies for managing failures and minimizing their impact. System logs generated by these devices serve as valuable repositories of historical trends and past failures. The use of machine learning techniques for failure prediction has become commonplace, enabling the extraction of insights from past data to anticipate future behavior patterns. Recently, large language models have demonstrated remarkable capabilities in tasks including summarization, reasoning, and event prediction. Therefore, in this paper, we endeavor to investigate the potential of large language models in predicting system failures, leveraging insights learned from past failure behavior to inform reasoning and decision-making processes effectively. Our approach involves leveraging data from the Intel® Computing Improvement Program (ICIP) system crash logs to identify significant events and develop CrashEventLLM. This model, built upon a large language model framework, serves as our foundation for crash event prediction. Specifically, our model utilizes historical data to forecast future crash events, informed by expert annotations. Additionally, it goes beyond mere prediction, offering insights into potential causes for each crash event. This work provides the preliminary insights into prompt-based large language models  for the log-based event prediction task.
\end{abstract}

\begin{IEEEkeywords}
log data, log analysis, large language model,
crash prediction using LLM, deep learning, machine learning.
\end{IEEEkeywords}
\begin{figure*}
\centering
\includegraphics[width=1\textwidth]{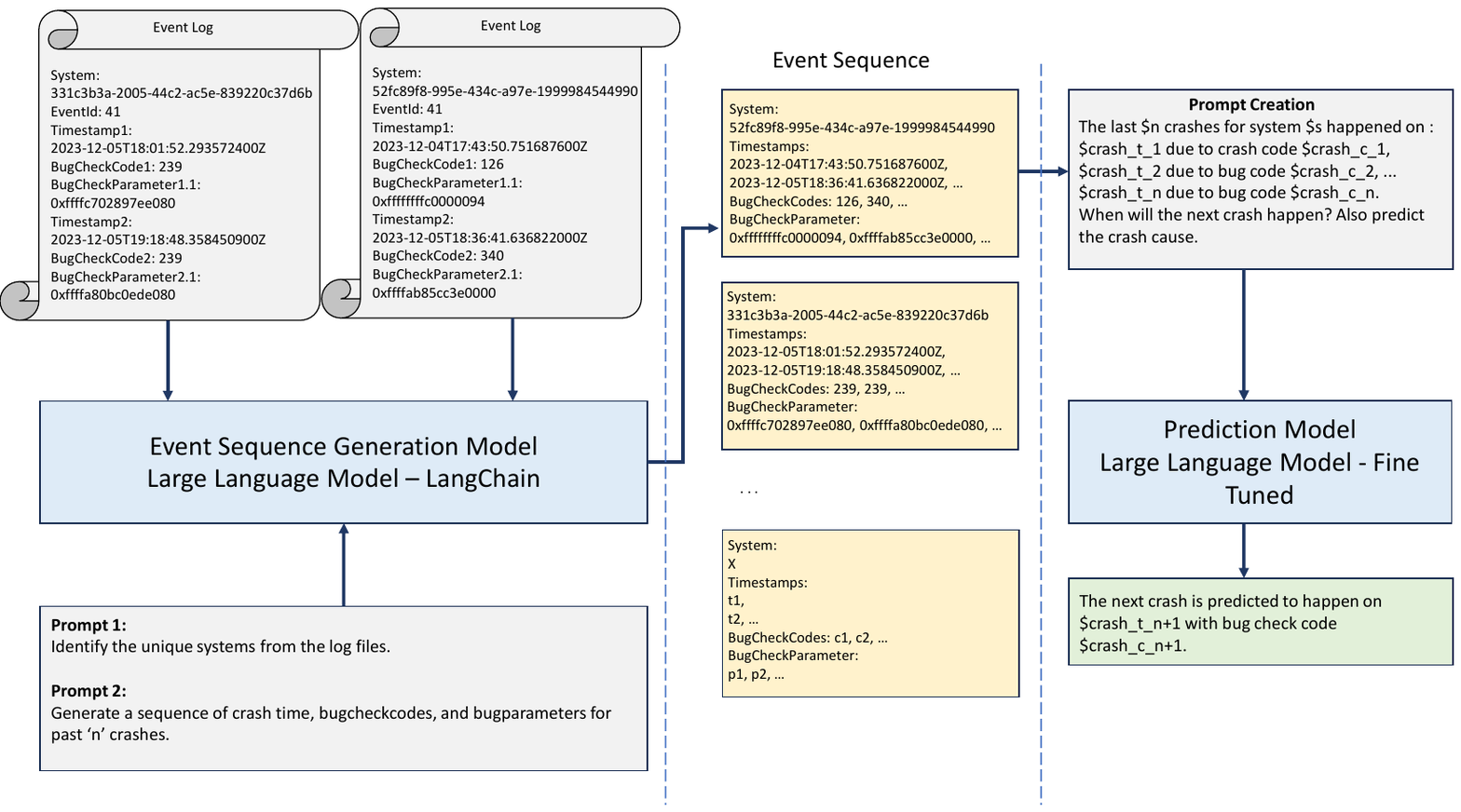}
\vspace{-1.0in}
\caption{\label{fig:architecture} Our framework utilizes advanced large language models to forecast forthcoming system crashes and their causes through a two-stage process. In the initial stage, an event sequence generation model sifts through raw event logs, identifying system crashes while capturing essential details such as timestamp, bug check code, and parameters. Following this, custom prompts are formulated to refine another large language model. This model undergoes extensive fine-tuning, scrutinizing historical events to detect patterns. Subsequently, it leverages this acquired insight to predict both the occurrence of the next crash event and its underlying cause.}
\end{figure*}

\section{Introduction}
Recent advancements in computing technology have caused a significant rise in our dependence on computing systems across a wide range of vital services. From banking operations, utility management, airlines, to information technology infrastructure, a plethora of everyday functions now rely on a diverse array of computing systems. The critical importance of minimizing system failures or downtime cannot be overstated, as it is often essential for societal function. System failures or crashes typically originate from system errors, where each critical error has the capacity to prompt system shutdowns or reboots in an effort to resolve the issues \cite{ref2}. Such errors can lead to considerable financial losses and inflict significant damage on crucial information technology (IT) infrastructure. Within the Windows environment, users often encounter the disruptive "blue screen of death" (BSOD), hampering device functionality \cite{ref2}. Recent such example is from crowdstrike incident \cite{crowdstrike}, where roughly 8.5 million systems crashed and were unable to properly restart in leading to the largest outage in the history of information technology. Typically, in the event of a system crash in the Windows operating system (OS), a crash log is generated to aid in the analysis of the crash triggers. Critical crashes are identified and logged as event 41. This event signifies that some unforeseen activity hindered the proper shutdown of Windows. Such shutdowns may result from power supply interruptions or stop errors. System crash logs serve as a comprehensive record of an IT system's operational activities and events, offering detailed insights. These logs are important in two folds: fault detection and failure prediction. Fault detection focuses on swiftly identifying the indicators of the critical failures as they arise, often employing anomaly detection methods. On the contrary, failure prediction adopts a proactive approach, aiming to issue early alerts regarding potential failures.

Several research work addressed the failure prediction problem based on machine learning and deep learning techniques \cite{7840733, 4470294}. A related study utilized telemetry data for system failure prediction through a concurrent work \cite{mudgal2024ensemblemethodfailuredetection}. Recent growth of large language models (LLMs) \cite{brown2020language, Radford2019LanguageMA, ouyang2022training} in the area of real world event-prediction and text-reasoning tasks have demonstrated interesting results \cite{shi2023language}. In particular, due to being pre-trained on text, logs, and code corpora, LLMs have also shown promising effectiveness in software engineering tasks, such as software comprehension \cite{ahmed2022fewshot, 9146834}, code generation and bug resolving \cite{du2023classeval, 10.1145/3533767.3534390, poesia2022synchromesh, du2023resolving}, log analysis \cite{10.1007/978-981-99-7587-7_13, liu2024interpretable, le2023log, jiang2024lilac, pan2023raglog, Ma_2024}, and failure mode classification \cite{stewart2023large}. Therefore, it becomes interesting to investigate their capability in system crash prediction. In this paper, we investigate their capabilities in predicting and reasoning about the system crash events given the past. Specifically, we model system's sequences of crash timestamps and predict their future crashes along with the possible cause of failure. Our hypothesis is that LLMs are potentially useful for advancing solutions to this problem because event sequences are often accompanied with rich text information which LLMs can handle.

This paper introduces CrashEventLLM, a novel framework designed to integrate a large language model (LLM) for event prediction. The framework, depicted in Fig. \ref{fig:architecture}, is based on the utilization of crash log data obtained from the Intel® Computing Improvement Program (ICIP) \cite{7936346}, focusing on a diverse range of client devices and containing crucial crash information such as timestamps and bugcheckcodes. The guid in Fig. \ref{fig:architecture} have been altered to safeguard the privacy of the users of those systems. The framework organizes raw crash logs into sequences by leveraging a pretrained LLM, which are then utilized to fine-tune a second LLM for predicting future failure occurrences within a specified time window. This predictive model learns to perform abductive reasoning, informed by historical crash events within a defined time frame, to generate potential causes for each predicted crash. The generality of our prediction modeling framework allows for the incorporation of various large language models. Through our experiments with different model sizes, we demonstrate the competency of LLMs in crash event prediction, regardless of their scale, when fine-tuned for this specific task.

\section{\textbf{Method}}
In this section, we present an overview of both the dataset utilized and the models subjected to evaluation. Fig. \ref{fig:architecture} illustrates the framework of our architecture.

\subsection{\textbf{Dataset}}
The dataset comprises information gathered from systems whose users have opted to participate in data collection and analytics (DCA) using ICIP telemetry software \cite{7936346}. ICIP functions as a telemetry software tool for monitoring product health, provided to users when they visit www.intel.com for driver downloads. Primarily, this data includes client PCs featuring various generations of central processing units (CPUs). DCA retrieves data from machines exclusively during their operational phases, known as the S0 state, and collects it at regular intervals, typically every 5 seconds. On-device aggregation of data occurs every 24 hours, with the aggregated data uploaded to the datastore when the system is active and connected to the network. Data is accessible only for the days when the machine is active, specifically in the S0 state, for at least a few seconds. 

\subsection{\textbf{Large Language Models}}

Language models acquire knowledge through text comprehension. In recent years, LLMs trained on vast amounts of internet text have exhibited remarkable performance across a spectrum of challenging tasks, including arithmetic reasoning and multi-turn dialogue \cite{wei2023chainofthought, openai2024gpt4}. Recent research \cite{gruver2023large} has further highlighted the exceptional performance of LLMs on time-series or sequential data, attributed to their inherent capacity to represent multimodal distributions, which aligns well with salient features in many time series, such as recurring seasonal trends. Additionally, LLMs have demonstrated proficiency in log analytics tasks such as anomaly detection and log parsing \cite{liu2024interpretable, jiang2024lilac, pan2023raglog, Ma_2024}. However, as noted by Mudgal et al. \cite{10.1007/978-981-99-7587-7_13}, event prediction using LLMs with zero-shot learning has exhibited suboptimal performance. Therefore, an intriguing area for exploration is the evaluation of large language models for system crash prediction employing few-shot learning techniques. Given that logs comprise complex and critical information often requiring preprocessing, we devise an event sequence generation model based on retrieval augmented generation (RAG) architecture \cite{lewis2021retrievalaugmented} to extract crash events and generate event sequences featuring the timing and causes of previously recorded crashes. Subsequently, we construct prompts utilizing the generated crash sequence data to feed into the prediction model, which then forecasts future crashes. Both models are elucidated in Sections \ref{eventseq} and \ref{predictionmodel}.

\subsubsection{\textbf{Event sequence generation model}} \label{eventseq} 
The objective of this module is to produce sequences of crash events in the form ($t_1$, $k_1$), ($t_2$, $k_2$), . . ., where 0 $\textless$ t1 $\textless$ t2 $\textless$ . . . denote the times of crash occurrences in sequential manner, and each $K_i$ $\in$ K represents a discrete crash type. Our aim is to create linearized sequences encapsulating crash event concepts. Initially, we present the raw log files to a LLM and guide it through a series of prompts to identify unique systems within the logs and subsequently discern the crashes associated with them by using langchain. Formally, given a raw crash log file $L$, we task the LLM with identifying $N$ unique systems and generating a sequence of the past $M$ crash events arranged in chronological order as a pair of time and reason of crash ($t_i$, $k_i$). This process facilitates the extraction of structured information from the raw crash logs, encompassing system details, crash timestamps, and reasons for the crashes. Following extraction, the information is partitioned into distinct segments using a fixed-size time window, resulting in two types of sequences: a time sequence detailing the timestamps of past crash events while excluding irrelevant content, and a crash cause sequence providing a high-level abstraction of sequential crash patterns to support pattern-based event prediction.

\begin{figure*}
\setkeys{Gin}{width=.33\linewidth} 
\begin{minipage}[t]{2\columnwidth}
  {\label{fig:time}\includegraphics{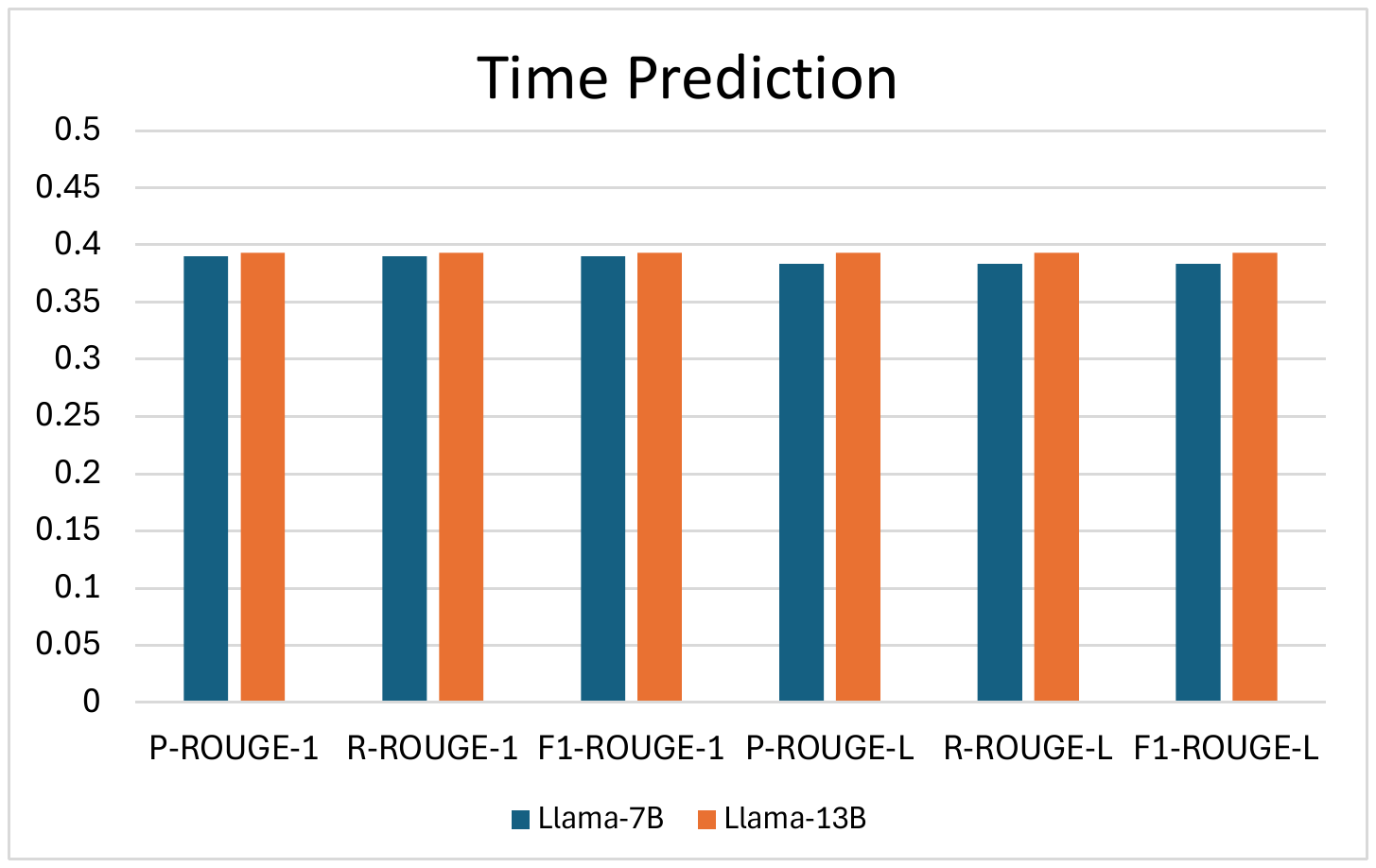}} \hfill
  {\label{fig:cause}\includegraphics{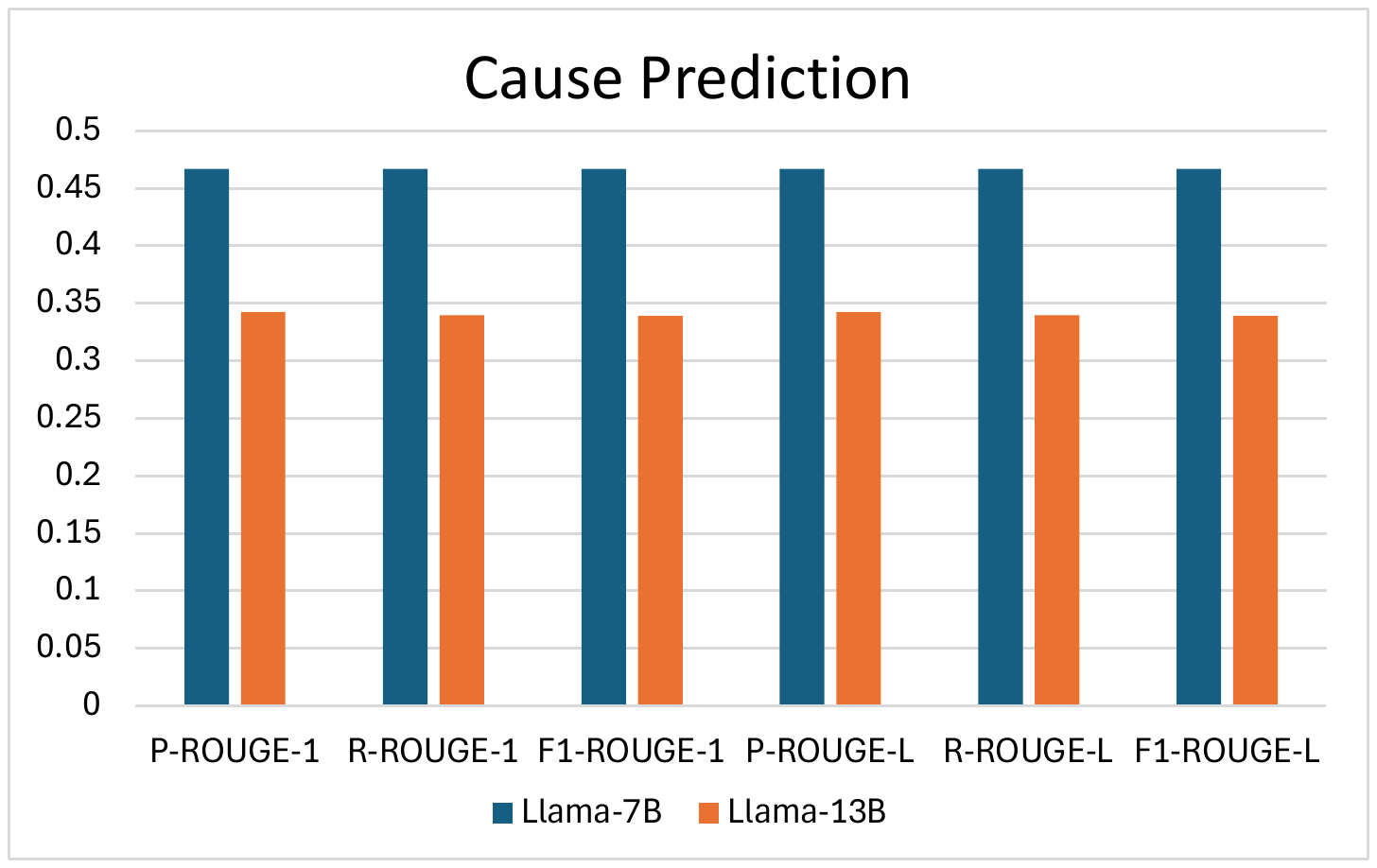}}\hfill
  {\label{fig:full}\includegraphics{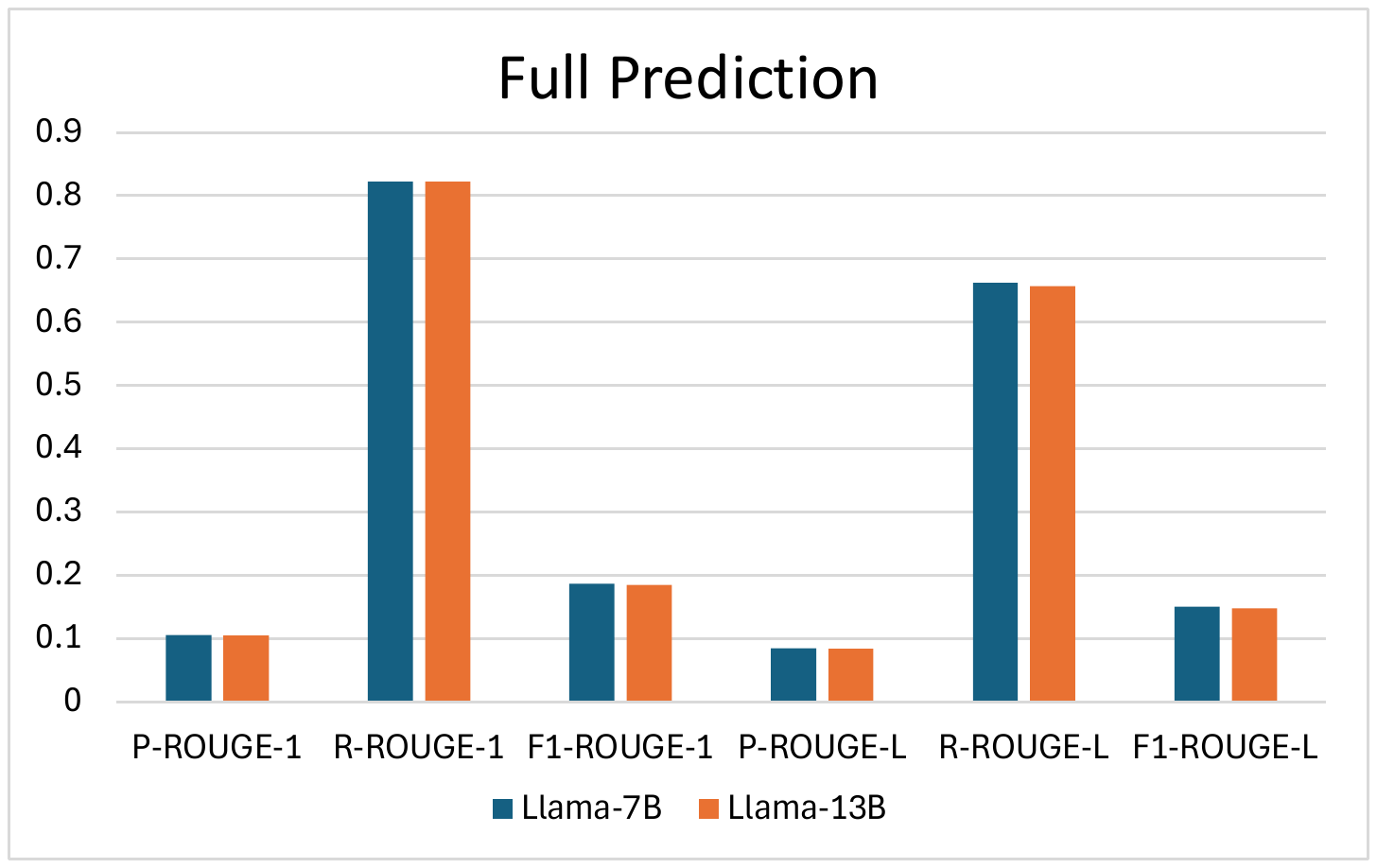}}\hfill
  \caption{ Prediction performance of Llama2-7B and Llama2-13B. The left figure shows the ROUGE scores for crash time prediction, the middle figure for crash cause prediction, and the right figure for full prediction. While both LLMs perform comparably in time and full prediction, surprisingly, Llama2-7B demonstrates superior performance in crash cause prediction.} \label{fig:results}
\end{minipage}\hfill 

\end{figure*}

\subsubsection{\textbf{Prediction model}} \label{predictionmodel} The goal of this module is to predict the next event for a given history of events $H_i$ = ($t_1$, $k_1$), ($t_2$, $k_2$), $\cdot \cdot \cdot $, ($t_{i - 1}$, $k_{i - 1}$). Precisely, it consists of two subtasks: the first is to predict the time $t_i$ of the next event; the second is to predict the type $k_i$ of the next event with the knowledge of its time $t_i$. The standard approach is to build a probabilistic model over the sequences. Such models typically define an intensity function $\lambda_k$: the intensity value $\lambda_k$$(t)$ is the instantaneous rate that an event of type $k$ occurs at time $t$. Given the function $\lambda_k$, one could obtain the minimum Bayes risk (MBR) prediction of the next event given the history \cite{https://doi.org/10.1002/nav.3800260304}. Specifically, we start with a set of past crash events $E_t$ for a randomly selected system $X$ in the form - “When will the next crash happen on system $X$?”. Based on the model prediction, we append additional question to predict the cause of crash given the past crash causes $E_c$ in the form - "What will be the predicted crash cause? This LLM fundamentally utilizes the sequence of crash events that was produced by event sequence generation model to predict the occurrence of the next crash.

Overall, as shown in the architecture diagram Fig. \ref{fig:architecture}, our design serves dual purposes: (i) preprocesses the logs to extract the information of crash events (ii) leverages in-context learning of the past crashes and predicts the future crashes with the possible cause.

\subsection{\textbf{Experimental Setup}}
We conduct experiments utilizing two LLMs: Llama-2, possessing 7 billion parameters, and another version of it with 13 billion parameters, as described by Touvron et al. \cite{touvron2023llama}, referred to as Llama2-7B and Llama2-13B, respectively. For our dataset, we employed 10-shot prompts, where each \enquote{shot} consists of a demonstration comprising an effect event followed by one or more expert-annotated cause events. Both models underwent fine-tuning on an annotated dataset comprising 100 (observation, label) pairs and were subsequently validated on a 40-pair validation set.

\section{\textbf{Evaluation Results}}

\subsection{\textbf{Results}} Our main results are demonstrated in Fig. \ref{fig:results}. We evaluate our results in three categories: time prediction, cause prediction, and full prediction. As the output of our prediction model is in natural language as a sentence with the predicted time and cause of crash, we postprocess the output to extract those information pieces and compute the ROUGE scores \cite{lin-2004-rouge, ganesan2018rouge}. ROUGE compares n-gram overlap of words on a surface level. We chose ROUGE-1 and ROUGE-L as they are the most commonly annotated ROUGE variants used for evaluating text generation. In ROUGE-1 and ROUGE-L, we show the evaluation results for precision, recall, and F1 for all three categories. The fine-tuned model predicted the output in a defined format for full validation set.

\subsection{\textbf{Analysis}} 

The Fig. \ref{fig:results} displays the outcomes obtained from our framework's Llama-7B and Llama-13B versions. Notably, the performance of the Llama-7B version rivals with Llama-13B version in both time prediction and full prediction tasks. However, the Llama-7B model surpasses the Llama-13B model in crash cause prediction. It's noteworthy that despite the significant parameter difference between Llama-7B and Llama-13B, the former outperforms the latter. To get the deep insights into this observation, it would be valuable to assess our approach on a larger validation set, given the relatively small size of our current validation dataset. Furthermore, we acknowledge the potential for enhancing our result evaluation through improved postprocessing techniques. For instance, previous studies have indicated that removing stopwords from generated output can lead to improved ROUGE scores \cite{ganesan2018rouge}. Therefore, we intend to explore such techniques for refining our evaluation processes.

The formulation of prompts holds substantial impact over the generated text's quality. Despite this, our methodology involved crafting prompt templates without preliminary experimentation or customization tailored to our dataset. Instead, we conducted post-analysis by assessing various templates on a small subset of our validation data. Surprisingly, we noted minimal discrepancies in results, provided the task description was clearly outlined, which was also highlighted in previous research work \cite{shi2023language}.

\section{\textbf{Limitations and Future Directions}}
Our framework utilizes LLMs like Llama-7B and Llama-13B for generating event sequences and predicting crash events. However, it inherits potential limitations of these models, such as issues related to hallucination and biased content. Consequently, there's a risk of generating irrelevant, inaccurate, or misleading crash events, thereby compromising our framework's overall performance. Moreover, our dataset comprises crash events from heterogeneous systems with diverse characteristics, potentially leading to distinct crash patterns. Due to the small size of our dataset, LLMs might struggle to learn effectively, resulting in poor ROUGE scores.

For future research in system crash prediction, we propose three directions: First, fine-tuning more advanced LLMs with specific data and conducting large-scale studies could unlock new arenas for system crash prediction using crash logs and LLMs. Second, expanding the dataset size may enhance performance. Specifically, Mudgal et al. \cite{mudgal2024ensemblemethodfailuredetection} utilized system telemetry data to detect system failures. Similarly, integrating telemetry data including system characteristics and software updates with system crash information could enhance the identification of correlations between telemetry data and system crashes, leading to more accurate predictions of system failures or crashes. Third, while we evaluated predicted outputs using ROUGE scores in this paper, advancements in LLMs warrant exploring additional performance metrics for evaluation purposes.

\section{\textbf{Conclusion}}

In this paper, we investigate the potential of LLMs in system crash prediction by using the system crash logs. We introduce the CrashEventLLM framework designed to leverage LLMs for crash event sequence generation and crash event prediction. Through experiments and model fine-tuning, we demonstrate the competency of LLMs in predicting crash events, offering insights into potential causes. The versatility of our prediction modeling framework allows for the integration of various LLMs, highlighting the adaptability of LLMs in addressing specific tasks like crash event prediction. The implications of our findings for future research in crash event sequence generation and crash event predictions are considerable, suggesting areas for potentially impactful innovations with further exploration in this field.

\section{\textbf{Acknowledgement}}
We gratefully acknowledge the contributions of our colleague Joshua Boelter.

\bibliographystyle{IEEEtran}
\bibliography{ipbib}

\end{document}